\documentstyle[axodraw]{article}

\begin{document}
\vspace*{-1in}
\renewcommand{\thefootnote}{\fnsymbol{footnote}}
\begin{flushright}
SINP/TNP/99-25\\
\texttt{hep-ph/9907432} 
\end{flushright}
\vskip 5pt
\begin{center}
{\Large{\bf Neutrino mass and magnetic moment in supersymmetry without
$R$-parity in the light of recent data}}
\vskip 25pt
{\sf Gautam Bhattacharyya $^{1,2}$}, 
{\sf H.V. Klapdor-Kleingrothaus $^{1}$},
{\sf Heinrich P\"{a}s $^{1}$}
\vskip 10pt
$^1$ {\small Max-Planck-Institut f\"{u}r Kernphysik, P.O. Box 103980,
D-69029 Heidelberg, Germany} \\
$^2$ {\small Saha Institute of Nuclear Physics, 1/AF Bidhan Nagar, 
Calcutta 700064, India} 

\vskip 20pt

{\bf Abstract}
\end{center}

\begin{quotation}
{\small We consider the generation of neutrino Majorana mass and
transition magnetic moment by the lepton-number violating $\lambda$
and/or $\lambda'$ couplings in $R$-parity-violating supersymmetric
models. We update (and improve) the existing upper limits on the
relevant couplings using the most recent data on neutrino masses and
mixings, indicating also the possible improvement by the GENIUS
project. We study the implication of this update on the induced
neutrino magnetic moment.}
\end{quotation}

\vskip 20pt  

\setcounter{footnote}{0}
\renewcommand{\thefootnote}{\arabic{footnote}}
%\vfill
%\clearpage
%\setcounter{page}{1}
%\pagestyle{plain}

Supersymmetry without $R$-parity \cite{rpar} provides an elegant
mechanism for generating neutrino (Majorana) masses and mixings. In
these models, there are mainly two sources of neutrino mass
generation. In one scenario, products of trilinear $\lambda$ and/or
$\lambda'$ couplings generate a complete neutrino mass matrix through
one-loop self-energy graphs \cite{trilinear,recent}. In this framework,
the diagrams that generate the masses also generate magnetic moments
by insertion of photons to the internal lines. In the other scenario,
the bilinear $R$-parity-violating (RPV) terms induce sneutrino vacuum
expectation values (VEV's) allowing neutrinos to mix with the
neutralinos. In this mechanism only one physical neutrino becomes
massive \cite{bilinear}. In the present analysis, although we
concentrate mainly on the trilinear $L$-violating couplings, we also
comment on the possible impact of the bilinear parameters. We use the
recent data on atmospheric \cite{sk} and solar \cite{solar} neutrinos,
the measurement of $\nu_e$ mass in the Troitsk \cite{moscow} and Mainz
\cite{mainz} tritium beta-decay experiments, and the measurement of
the effective neutrino mass $\langle m_\nu \rangle$ in the Heidelberg
- Moscow ${}^{76}$Ge experiment \cite{hm}, to update (and improve) the
existing limits \cite{review} on many different combinations of the
trilinear couplings from their contribution to neutrino masses. We
also calculate the magnetic moment of the neutrino, which has an
intimate connection to its mass (for previous studies of neutrino
magnetic moments in RPV models, see \cite{babu_moha,barbieri}).

Generally two types of magnetic moment may arise: (i) the Dirac-type
magnetic moment that rotates $\nu_{eL}$ to $\nu_{eR}$, and (ii) the
transition magnetic moment that takes $\nu_{eL}$ to $\nu^c_{\mu
R}$. If one adds a right-handed singlet neutrino to the standard model
(SM), a non-zero Dirac mass of the neutrino implies a non-zero
Dirac-type magnetic moment, given by \cite{fuji}
\begin{equation}
\mu_\nu = {{3 e G_F m_\nu}\over{8\pi^2\sqrt{2}}} = 
3 \times 10^{-19} \left({m_\nu}\over{{\rm 1~eV}}\right) \mu_B,
\end{equation}
where $\mu_B = e/2 m_e$ is the Bohr magneton. The supersymmetric
contribution to the Dirac-type magnetic moment could at most be
comparable to the above value \cite{ng}. On the other hand, a Majorana
neutrino, massive or massless, cannot possess a diagonal magnetic
moment due to the $CPT$ theorem. It may only have a transition
magnetic moment between two species having opposite $CP$
eigenvalues. In supersymmetric models without a right-handed neutrino,
the RPV couplings can generate only $\Delta L = 2$ Majorana
masses. The magnetic moment they induce is therefore only
transitional.

The $L$-violating part of the RPV superpotential can be
written as 
\begin{equation}
    {\cal W}_{RPV}  =  {1\over 2}\lambda_{ijk} L_i L_j E^c_k
                        +  \lambda'_{ijk} L_i Q_j D^c_k + \mu_i L_i H_u, 
\label{superpot}
\end{equation}
where $i$, $j$ and $k$ are quark and lepton generation indices. In
Eq.~(\ref{superpot}), $L_i$ and $Q_i $ are SU(2)-doublet lepton and
quark superfields, $E^c_i$ and $D^c_i$ are SU(2)-singlet charged
lepton and down-quark superfields, and $H_u$ is the Higgs superfield
responsible for the mass generation of the up-type quarks,
respectively. There are 9 $\lambda$-type (due to an antisymmetry in
the first two generation indices), 27 $\lambda'$-type and 3 $\mu_i$
couplings. Stringent upper limits exist on all these couplings from
different experiments \cite{review,faessler}.

We first consider the effects of the $\lambda'$ interactions. The
relevant part of the Lagrangian can be written as 
\begin{equation}
 - {\cal L}_{\lambda'} = \lambda'_{ijk} \left[\bar{d}_k P_L \nu_i
  \tilde{d}_{jL} + \bar{\nu}^c_i P_L d_j \tilde{d}^*_{kR}\right]
+ ~{\rm h.c.}   
\label{lagrangian}
\end{equation}
Majorana mass terms for the left-handed neutrinos, given by ${\cal
L}_M = -\frac{1}{2} m_{\nu_{ii'}} \bar{\nu}_{Li} \nu^c_{Ri'} +~{\rm
h.c.}$, are generated at one loop. Figs.~1(a) and 1(b) show the
corresponding diagrams. The induced masses are given by
\begin{equation} 
m_{\nu_{ii'}} \simeq {{N_c \lambda'_{ijk} \lambda'_{i'kj}}
\over{16\pi^2}} m_{d_j} m_{d_k}
\left[\frac{f(m^2_{d_j}/m^2_{\tilde{d}_k})} {m_{\tilde{d}_k}} +
\frac{f(m^2_{d_k}/m^2_{\tilde{d}_j})} {m_{\tilde{d}_j}}\right],
\label{mass}
\end{equation}     
where $f(x) = (x\ln x-x+1)/(x-1)^2$. Here, $m_{d_i}$ is the down quark
mass of the $i$th generation inside the loop, $m_{\tilde{d}_i}$ is an
average of $\tilde{d}_{Li}$ and $\tilde{d}_{Ri}$ squark masses, and
$N_c = 3$ is the colour factor. In deriving Eq.~(\ref{mass}), we
assumed that the left-right squark mixing terms in the soft part of
the Lagrangian are diagonal in their physical basis and proportional
to the corresponding quark masses, {\em i.e.} $\Delta m^2_{\rm LR} (i)
= m_{d_i} m_{\tilde{d}_i}$. At this point, a remark about the quark
mixing angles is in order. These angles are supposed to appear in the
loop amplitudes under consideration. The reason behind this is that
the RPV couplings have been written in the flavour basis, while the
states that propagate inside the loop are in their mass
basis. However, we have ignored this small difference in order not to
complicate the discussion unnecessarily. On the other hand, the
neutrinos may have large mixings and so the difference between their
flavour and mass eigenstates cannot be ignored.

Now one can study the electromagnetic properties of the neutrino,
namely the magnetic moment, by attaching photons to the internal lines
of the loops in Figs.~1(a) and 1(b). One can express the effective
Hamiltonian as 
\begin{equation}
{\cal H}_{\rm eff} = \mu_{\nu_{ii'}} \bar{\nu}_{Li}
\left(\frac{\sigma^{\alpha \beta}}{2}\right) \nu^c_{Ri'} F_{\beta
\alpha},
\end{equation}
where $\mu_{\nu_{ii'}}$ is the transition magnetic moment between the
states $\nu_i$ and $\nu_{i'}$. It is given by
\begin{equation}
\mu_{\nu_{ii'}} \simeq \left(1 - \delta_{ii'}\right)
{{N_c \lambda'_{ijk} \lambda'_{i'kj}}
\over{4\pi^2}} m_e Q_d m_{d_j} m_{d_k}
\left[\frac{g(m^2_{d_j}/m^2_{\tilde{d}_k})} {m^3_{\tilde{d}_k}} -
\frac{g(m^2_{d_k}/m^2_{\tilde{d}_j})} {m^3_{\tilde{d}_j}}\right]\mu_B,
\label{moment}
\end{equation} 
where $g(x) = (1+x)\ln x/(x-1)^3 - 2/(x-1)^2$, and the
$(1-\delta_{ii'})$ factor indicates that for $i = i'$ the magnetic
moment is zero. It is interesting to note that for $j = k$, even if
there is a contribution to the neutrino mass, the transitional
magnetic moment vanishes.

Since all down-type quarks are much lighter than the squarks, $x \ll
1$ (for all generations) is a good approximation. This implies $f(x)
\simeq 1$, and $g(x) \simeq -\ln x - 2$.  We now consider the
following two cases. In the first case, the down squarks of $j$th and
$k$th generation are degenerate, {\em i.e.} $m_{\tilde{d}_j} =
m_{\tilde{d}_k} \equiv \tilde{m}$. Then
\begin{equation}
\mu_{\nu_{ii'}} \simeq \left(1 - \delta_{ii'}\right)
8 m_{\nu_{ii'}} \frac{m_e Q_d}{\tilde{m}^2}
\ln \left(\frac{m_{d_j}}{m_{d_k}}\right) ~\mu_B.
\label{degenerate}
\end{equation}
In the second case, the down squark of the $j$th (say) generation is
much heavier than that of the $k$th generation; so only one term
within the square bracket of Eq.~(\ref{moment}) effectively
contributes. Then
\begin{equation}
\mu_{\nu_{ii'}} \simeq \left(1 - \delta_{ii'}\right)
8 m_{\nu_{ii'}} \frac{m_e Q_d}
{m^2_{\tilde{d}_k}} \left[\ln
\left(\frac{m_{\tilde{d}_k}}{m_{d_j}}\right) - 1\right] ~\mu_B.
\label{non_degenerate}
\end{equation}

With $\lambda$-type interactions, one obtains exactly similar results
as in Eqs.~(\ref{lagrangian}) to (\ref{non_degenerate}). The quarks
and squarks in these equations will be replaced by the leptons and
sleptons of the corresponding generations. The colour factor $N_c = 3$
in Eqs.~(\ref{mass}) and (\ref{moment}) and $Q_d$ in
Eqs.~(\ref{moment}) to (\ref{non_degenerate}) would be replaced by $1$
and $Q_e$ respectively. We do not explicitly write them down. We draw
the attention of the readers to the fact that Barbieri {\em et al.}
\cite{barbieri} have overlooked a part of the Lagrangian associated
with a given pair of $\lambda$ while presenting their analytic
formulae and the corresponding estimates. Since they assumed that
sleptons of different generations are fairly degenerate, (a) their
expression of the neutrino mass [Eq.~(3a) of \cite{barbieri}] should
be multiplied by a factor of 2, and (b) their expressions of the
magnetic moment [Eqs.~(3b) and (4) of \cite{barbieri}, which are
similar to our Eq.~(\ref{non_degenerate})] should correspond to our
Eq.~(\ref{degenerate}). Indeed these details do not change the
order-of-magnitude estimates they have presented.

In Table 1, we have presented the upper limits on different product
couplings calculated from the most recent bounds on the different
entries of the neutrino mass matrix (in flavour space). We have also
presented the transition magnetic moments between different flavours
derived using the most stringent constraints on the relevant product
couplings. Among the different entries of the flavour space mass
matrix, only the $ee$-term has a {\em direct experimental} bound
$m_{ee} < 0.2$ eV obtained by the Heidelberg - Moscow neutrinoless
double-beta decay experiment \cite{hm}. The planned GENIUS project in
the 1 and 10 ton versions \cite{genius} would be sensitive to $m_{ee}$
as low as 0.01 and 0.001 eV respectively. We have displayed their
possible impact in Table 1. To constrain the other entries of the mass
matrix we use (a) the limit on the lightest neutrino mass eigenstate
$m_1 < 2.5$ eV obtained by the Troitsk tritium beta-decay experiment
\cite{moscow}, in conjunction with (b) the oscillation solutions of
the atmospheric and solar neutrino problems \cite{sk,solar}, implying
\begin{equation}
\Delta m_{12} = \Delta m_{\odot}\ll \Delta m_{13}= 
\Delta m_{atm}
\ll {\cal O} 
(1 {\rm eV})< m_1,
\end{equation} 
in a three neutrino framework. Hence from unitarity an upper bound
$m_{\nu_{ii'}}< 2.5$ eV can be obtained for any flavour combination
$(i, i')$. A larger spectrometer planned for tritium beta-decay
experiment \cite{bonn} would bring the limit down to $m_{\nu_{ii'}} <
1$ eV. While most of the existing bounds are simply the products of
the bounds on the individual couplings given in \cite{wein}, there are
some combinations which receive stringent limits from $\mu e$
conversion in nuclei \cite{huitu}. The chirality flips in our
Figs. 1(a) and 1(b) explain why with heavier fermions inside the loop
the bounds are tighter. For this reason, we have presented the bounds
only for $j, k= 2, 3$.  In the process, constraints on the individual
couplings $\lambda_{122}$, $\lambda_{133}$, $\lambda_{233}$,
$\lambda_{322}$, $\lambda'_{i33}$ and $\lambda'_{i22}$ (for all $i$)
have improved quite significantly.  The most stringent upper limits
are then employed to derive the maximum magnetic moment that RPV
models can induce, using Eq.~(\ref{degenerate}) for the degenerate
case and Eq.~(\ref{non_degenerate}) for the hierarchical case. Since
Eq.~(\ref{non_degenerate}) is asymmetric in $j$ and $k$, both cases
$m_{\tilde{d}_j}\gg m_{\tilde{d}_k}$ and $m_{\tilde{d}_k}\gg
m_{\tilde{d}_j}$ are separately examined. For numerical purpose, we
have assumed the mass of whatever scalar is relevant to be 100 GeV
throughout, to be consistent with common practice and, in particular,
to compare with the old bounds. While for sleptons this sounds a
reasonable approximation, for squarks the present lower limit, even in
RPV scenario, is around 250 GeV \cite{d0}. In any case, for different
squark masses one can easily derive the appropriate bounds by
straightforward scaling. It should be noted that the product couplings
under consideration contribute to charged lepton masses as well. But
with the present limits those contributions are too small to be of any
relevance.

The bilinear couplings (the last term of Eq.~(\ref{superpot})) could
be another source of neutrino mass generation. Only one of the
neutrino physical states becomes massive if $\mu_\alpha$ and $\langle
L_\alpha \rangle$ (where, $\alpha = 0,~i$) are not aligned
\cite{bilinear}. Here, $\mu_0 \equiv \mu$ (note, $\mu H_d H_u$ appears
in the $R$-parity-conserving superpotential) and $L_0 \equiv H_d$ (the
Higgs superfield responsible for down quark mass generation). The
misalignment induces sneutrino VEV allowing neutrinos to mix with the
neutralinos, and, as a result, a neutrino mass is generated. The
implications of the bilinear couplings in the light of
Super-Kamiokande (SK) data and neutrinoless double-beta decay have
been studied in Ref.~\cite{faessler} (in a restricted scenario) and
Ref.~\cite{hirsch}, respectively. Since the neutrino mass is induced
at tree level, this interaction cannot generate any neutrino magnetic
moment. In this context, an interesting idea in the framework of
supergravity models has been proposed in Ref.~\cite{joshipura}. The
presence of trilinear RPV interaction at a high scale induces bilinear
RPV terms in the effective potential at the weak scale. The sneutrino
VEV induced this way causes neutrino-neutralino mixing. The neutrino
mass generated through this mixing dominates over the loop mass (due
to the presence of a large logarithm). The constraints on
$\lambda'_{ikk}$ obtained in this specific scheme are somewhat
stronger (barring accidental cancellation) than obtained from direct
loop induced processes.

At this point, a few comments on the existing bounds on neutrino
magnetic moment are in order. From SK solar neutrino data, a limit
$\mu_{\nu_e} < 1.6 \times 10^{-10}~\mu_B$ has recently been obtained
\cite{beacom}.  From SN1987A a bound $\mu_{\nu_e} < 10^{-10}~\mu_B$
was obtained \cite{sn}, although the bound from supernova applies only
to Dirac-type magnetic moments. The present experimental upper limits
on $\mu_{\nu_e} < 1.8 \times 10^{-10}~\mu_B$ and $\mu_{\nu_\mu} < 7.4
\times 10^{-10}~\mu_B$ have been obtained from neutral current
scattering data \cite{pdg,kim}. For the tau neutrino, a new bound
$\mu_{\nu_{\tau}} < 1.3 \times 10^{-7}~\mu_B$ has recently been
obtained from SK data \cite{gninenko}. In the near future, a new
detector of the MUNU Collaboration \cite{munu} installed near one of
the reactors of the Bugey power station should be sensitive to
transition magnetic moments (relating $\nu_e$ to other states) of
order $(2-3) \times 10^{-11}~\mu_B$.

Since a large neutrino magnetic moment ($10^{-10}-10^{-11}$ $\mu_B$)
can provide an elegant solution to the solar neutrino problem
\cite{okun}, by flipping $\nu_e$ to undetectable species, it has
always been a theoretically challenging exercise \cite{largemunu} to
build models which can generate a sizable magnetic moment of the
neutrino, keeping its mass small. But unless some additional symmetry
is invoked, it is difficult to endow a neutrino with a large magnetic
moment, since the same diagrams that induce neutrino masses also
generate magnetic moments (by insertion of photons to internal
lines). The task therefore is to look for some continuous or discrete
horizontal symmetry that would allow the magnetic moment but suppress
the mass of the neutrino. As an example, Voloshin \cite{voloshin}
considered an SU(2) horizontal symmetry, acting on the $(\nu, \nu^c)$
doublet, under which the neutrino Majorana mass term transforms as a
triplet and is therefore prohibited in the limit of exact symmetry;
the neutrino magnetic moment, on the other hand, transforms as a
singlet and hence can be large. In the context of RPV models, imposing
a horizontal symmetry was shown to provide at most an order of
magnitude enhancement \cite{barbieri}. Since the limits on RPV
couplings induced magnetic moment we obtain in the present analysis,
are at least a few orders of magnitude below the present level of
experimental sensitivity and the requirement for the solar neutrino
solution, it is unlikely that any additional symmetry imposed on the
RPV Lagrangian could provide the necessary enhancement to close the
gap.

To conclude, we have improved the upper limits on many individual and
product couplings of the $\lambda$- and $\lambda'$-types, from their
contribution to neutrino masses, using the most recent data. We have
also indicated how these limits will be further strengthened once the
planned experiments are realized. Attaching photons to the internal
lines of the loop diagrams that generate neutrino masses also induces
transition magnetic moments. The maximum induced magnetic moments we
obtain are too small to be tested in the present laboratory
experiments.

\vspace{5pt}
\noindent G.B. acknowledges warm hospitality at the
Max-Planck-Institut f\"{u}r Kernphysik, Heidelberg.

\begin{table}
\small
\caption{{\small Correlation among neutrino mass bounds, upper limits
on RPV couplings and neutrino magnetic moments. We have used $m_d$=9
MeV, $m_s$= 170 MeV, $m_b$=4.4 GeV \protect{\cite{pdg}}. The potential
of the planned GENIUS project in the 1 and 10 ton versions is also
exhibited. If not mentioned, the previous bounds are products of
bounds on individual couplings given in \protect{\cite{wein}}. For
$\lambda$-products, $m_{\tilde{d}}$ should be read as
$m_{\tilde{e}}$. The relevant scalars are always assumed to have a
common mass of 100 GeV.}}
\begin{tabular}{cccccc}
\hline
\hline

$\lambda^{(')}_{ijk}\lambda^{(')}_{i^{'}kj}$
&                     Our &  Previous  &  
$\mu_{\nu_{ii^{'}}}$
& $\mu_{\nu_{ii^{'}}}$  
& $\mu_{\nu_{ii^{'}}}$\\

& Bounds & Bounds &$(m_{\tilde{d}_j}=m_{\tilde{d}_k})$
&$(m_{\tilde{d}_j}\gg m_{\tilde{d}_k})$
&$(m_{\tilde{d}_k}\gg m_{\tilde{d}_j})$\\
\hline
\hline

$m_{ee}<0.2$ eV &&&& \\ \hline

$\lambda^{'}_{133}\lambda^{'}_{133}$ & $3.0 \cdot 10^{-8}$ & $4.9
\cdot 10^{-7}$ & 0& 0 & 0\\

$\lambda^{'}_{132}\lambda^{'}_{123}$ & $7.5 \cdot 10^{-7}$ & $1.6
\cdot 10^{-2}$ & 0& 0 & 0 \\

$\lambda^{'}_{122}\lambda^{'}_{122}$ & $1.8 \cdot 10^{-5}$ & $4.0
\cdot 10^{-4}$ & 0& 0 & 0\\

$\lambda_{133}\lambda_{133}$ & $5.3 \cdot 10^{-7}$ & $9.0 \cdot
10^{-6}$ & 0& 0 & 0\\

$\lambda_{132}\lambda_{123}$ & $8.7 \cdot 10^{-6}$ & $2.0 \cdot
10^{-3} $ & 0& 0 & 0\\

$\lambda_{122}\lambda_{122}$ & $1.4 \cdot 10^{-4}$ & $1.6 \cdot
10^{-3}$ & 0& 0 & 0\\ \hline

$m_{ee}<0.01 (0.001)$ eV & [GENIUS 1(10)t] &&& \\ \hline

$\lambda^{'}_{133}\lambda^{'}_{133}$ & $1.5 \cdot 10^{-9(-10)}$ & $4.9
\cdot 10^{-7}$ & 0 & 0 & 0\\

$\lambda^{'}_{132}\lambda^{'}_{123}$ & $3.7 \cdot 10^{-8(-9)}$ & $1.6
\cdot 10^{-2}$ & 0& 0 & 0\\

$\lambda^{'}_{122}\lambda^{'}_{122}$ & $9.2 \cdot 10^{-7(-8)}$ & $4.0
\cdot 10^{-4}$ & 0& 0 & 0\\

$\lambda_{133}\lambda_{133}$ & $2.6 \cdot 10^{-8(-9)}$ & $9.0 \cdot
10^{-6}$ & 0& 0 & 0\\

$\lambda_{132}\lambda_{123}$ & $4.3 \cdot 10^{-7(-8)}$ & $2.0 \cdot
10^{-3} $ & 0& 0 & 0\\

$\lambda_{122}\lambda_{122}$ & $7.1 \cdot 10^{-6(-7)}$ & $1.6 \cdot
10^{-3}$ & 0& 0 & 0\\ \hline

$m_{e\mu}<2.5$ eV &&&&\\ \hline

$\lambda^{'}_{133}\lambda^{'}_{233}$ &  $3.8 \cdot 10^{-7}$ &   
$3.5 \cdot 10^{-5}$ \cite{huitu} & 
0    &    0    &      0 \\

$\lambda^{'}_{132}\lambda^{'}_{223}$ & $9.3 \cdot 10^{-6}$ & $2.4
            \cdot 10^{-2}$ & $1.1 \cdot 10^{-15}$ & $7.2 \cdot
            10^{-16}$ & $1.8 \cdot 10^{-15}$\\

$\lambda^{'}_{123}\lambda^{'}_{232}$ &  $9.3 \cdot 10^{-6}$ &   
$1.5 \cdot 10^{-2}$ &    $1.1 \cdot 10^{-15}$ & $1.8 \cdot 10^{-15}$      
& $7.2 \cdot 10^{-16}$ \\ 

$\lambda^{'}_{122}\lambda^{'}_{222}$ &  $2.3 \cdot 10^{-4}$ &   
$3.3 \cdot 10^{-7}$ \cite{huitu}  & 0   & 0  & 0 \\

$\lambda_{133}\lambda_{233}$ & $6.6 \cdot 10^{-6}$ & $1.7 \cdot
10^{-5}$ \cite{huitu} & $0$ & 0  & 0 \\ 

$\lambda_{123}\lambda_{232}$ & $1.1 \cdot 10^{-4}$ & $2.0 \cdot
10^{-3}$ & $2.9 \cdot 10^{-15}$ & $6.0 \cdot 10^{-15}$ & $3.1 \cdot
10^{-15}$ \\

\hline

$m_{\mu\mu}<2.5$ eV &&&&\\ \hline

$\lambda^{'}_{233}\lambda^{'}_{233}$ & $3.8 \cdot 10^{-7}$ & $1.4
\cdot 10^{-1}$ & 0 & 0 & 0\\ $\lambda^{'}_{232}\lambda^{'}_{223}$ &
$9.3 \cdot 10^{-6}$ & $2.2 \cdot 10^{-2}$ & 0 & 0 & 0\\

$\lambda^{'}_{222}\lambda^{'}_{222}$ & $2.3 \cdot 10^{-4}$ & $3.6
\cdot 10^{-3}$ & 0 & 0 & 0\\ $\lambda_{233}\lambda_{233}$ & $6.6 \cdot
10^{-6}$ & $2.5 \cdot 10^{-3}$ & 0 & 0 & 0\\ \hline

$m_{e\tau}<2.5$ eV &&&&\\ \hline

$\lambda^{'}_{133}\lambda^{'}_{333}$ & $3.8 \cdot 10^{-7}$ & $1.2
\cdot 10^{-4}$ & 0 & 0  & 0\\

$\lambda^{'}_{132}\lambda^{'}_{323}$ & $9.3 \cdot 10^{-6}$ & $1.4
\cdot 10^{-1}$ & $1.1 \cdot 10^{-15}$ & $7.2 \cdot 10^{-16}$ & $1.8
\cdot 10^{-15}$\\

$\lambda^{'}_{123}\lambda^{'}_{332}$ &  $9.3 \cdot 10^{-6}$ &    
$6.8 \cdot 10^{-3}$ &   $1.1 \cdot 10^{-15}$ &  $1.8 \cdot 10^{-15}$ & 
$7.2 \cdot 10^{-16}$ \\

$\lambda^{'}_{122}\lambda^{'}_{322}$ & $2.3 \cdot 10^{-4}$ & $7.2
\cdot 10^{-3}$ & 0 & 0  & 0 \\

$\lambda_{132}\lambda_{323}$ & $1.1 \cdot 10^{-4}$ & $2.5 \cdot
10^{-3}$ & $2.9 \cdot 10^{-15}$ & $3.1 \cdot 10^{-15}$ & $6.0 \cdot
10^{-15}$ \\

$\lambda_{122}\lambda_{322}$ & $1.8 \cdot 10^{-3}$ & $2.0 \cdot
10^{-3}$ & 0 & 0  & 0  \\ \hline

$m_{\mu\tau}<2.5$ eV &&&&\\ \hline

$\lambda^{'}_{233}\lambda^{'}_{333}$ & $3.8 \cdot 10^{-7}$ & $6.3
\cdot 10^{-2}$ & 0 & 0  & 0 \\

$\lambda^{'}_{232}\lambda^{'}_{323}$ & $9.3 \cdot 10^{-6}$ & $1.3
\cdot 10^{-1}$ & $1.1 \cdot 10^{-15}$ &$7.2 \cdot 10^{-16}$& $1.8
\cdot 10^{-15}$ \\

$\lambda^{'}_{223}\lambda^{'}_{332}$ & $9.3 \cdot 10^{-6}$ &    
$1.0 \cdot 10^{-2}$ &  $1.1 \cdot 10^{-15}$ &  $1.8 \cdot 10^{-15}$  &   
$7.2 \cdot 10^{-16}$ \\

$\lambda^{'}_{222}\lambda^{'}_{322}$ & $2.3 \cdot 10^{-4}$ & $2.2
\cdot 10^{-2}$ & 0 & 0 & 0 \\

$\lambda_{232}\lambda_{323}$ & $1.1 \cdot 10^{-4}$ & $2.5 \cdot
10^{-3}$ & $2.9 \cdot 10^{-15}$ &

$3.1 \cdot 10^{-15}$ & $6.0 \cdot 10^{-15}$ \\ \hline

$m_{\tau\tau}<2.5$ eV &&&&\\ \hline

$\lambda^{'}_{333}\lambda^{'}_{333}$ & $3.8 \cdot 10^{-7}$ & $2.9
      \cdot 10^{-2}$ & 0 & 0 & 0\\

$\lambda^{'}_{332}\lambda^{'}_{323}$ & $9.3 \cdot 10^{-6}$ & $6.1
         \cdot 10^{-2}$ & 0 & 0 & 0 \\

$\lambda^{'}_{322}\lambda^{'}_{322}$ & $2.3 \cdot 10^{-4}$ & $1.3
          \cdot 10^{-1}$ & 0 & 0 & 0 \\

$\lambda_{322}\lambda_{322}$ & $1.8 \cdot 10^{-3}$ & $2.5 \cdot
          10^{-3}$ & 0 & 0 & 0 \\ \hline \hline
\end{tabular}
\end{table}

\newpage
%\end{document}
\noindent
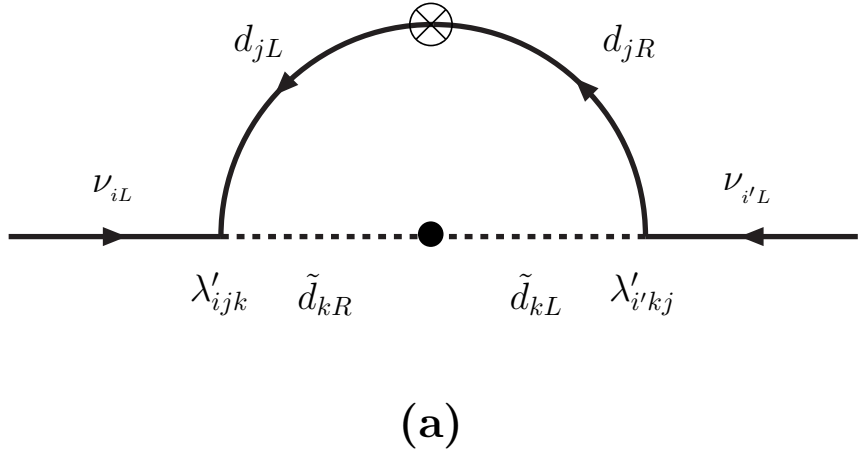
Figure 1: The $\lambda'$-induced one loop diagrams
contributing to Majorana masses for the neutrinos. To generate
magnetic moments photons should be attached to the internal lines.
\begin{center}
\begin{picture}(500,600)(0,0)
\SetWidth{2.0}
\ArrowLine(60,450)(140,450)
\ArrowArc(220,450)(80,0,90)
\ArrowArc(220,450)(80,90,180)
\DashLine(140,450)(220,450){3}
\DashLine(220,450)(300,450){3}
\ArrowLine(380,450)(300,450)
%\GCirc(180,452){4}{0}
\Text(220,452)[]{$\Huge{\boldmath{{\bullet}}}$}
\Text(220,532)[]{$\Huge{\boldmath{{\otimes}}}$}
\Text(100,470)[]{$\Large{\boldmath{\nu_{_{iL}}}}$}
\Text(340,470)[]{$\Large{\boldmath{\nu_{_{i'L}}}}$}
\Text(155,525)[]{$\Large{\boldmath{d_{jL}}}$}
\Text(295,525)[]{$\Large{\boldmath{d_{jR}}}$}
\Text(180,430)[]{$\Large{\boldmath{\tilde{d}_{kR}}}$}
\Text(260,430)[]{$\Large{\boldmath{\tilde{d}_{kL}}}$}
\Text(140,430)[]{$\Large{\boldmath{\lambda'_{ijk}}}$}
\Text(300,430)[]{$\Large{\boldmath{\lambda'_{i'kj}}}$}
\Text(220,380)[]{\LARGE{\bf{(a)}}}

\ArrowLine(60,150)(140,150)
\ArrowArcn(220,150)(80,90,0)
\ArrowArcn(220,150)(80,180,90)
\DashLine(140,150)(220,150){3}
\DashLine(220,150)(300,150){3}
\ArrowLine(380,150)(300,150)
%\GCirc(180,151){4}{0}
\Text(220,151)[]{$\Huge{\boldmath{{\bullet}}}$}
\Text(220,231)[]{$\Huge{\boldmath{{\otimes}}}$}
\Text(100,170)[]{$\Large{\boldmath{\nu_{_{iL}}}}$}
\Text(340,170)[]{$\Large{\boldmath{\nu_{_{i'L}}}}$}
\Text(155,225)[]{$\Large{\boldmath{d_{kR}}}$}
\Text(295,225)[]{$\Large{\boldmath{d_{kL}}}$}
\Text(180,130)[]{$\Large{\boldmath{\tilde{d}_{jL}}}$}
\Text(260,130)[]{$\Large{\boldmath{\tilde{d}_{jR}}}$}
\Text(140,130)[]{$\Large{\boldmath{\lambda'_{ijk}}}$}
\Text(300,130)[]{$\Large{\boldmath{\lambda'_{i'kj}}}$}
\Text(220,80)[]{\LARGE{\bf{(b)}}}

\end{picture}
\end{center}


\begin{thebibliography}{99}

\bibitem{rpar} G. Farrar, P. Fayet, Phys. Lett. 76B, 575 (1978); 
S. Weinberg, Phys. Rev. D 26, 287 (1982);
N. Sakai, T. Yanagida, Nucl. Phys. B 197, 533 (1982);
C. Aulakh, R. Mohapatra, Phys. Lett. 119B, 136 (1982).

\bibitem{trilinear} S. Dimopoulos, L. Hall, Phys. Lett. B207, 210
(1987); R. Godbole, P. Roy, X. Tata, Nucl. Phys. B 401, 67 (1993). 

\bibitem{recent} For recent studies, see M. Drees, S. Pakvasa, X. Tata
and T. ter Veldhuis, Phys. Rev. D 57, 5335 (1998); S. Rakshit,
G. Bhattacharyya, A. Raychaudhuri, Phys. Rev. D 59, 091701 (1999);
R. Adhikari, G. Omanovic, Phys. Rev. D 59, 073003 (1999); O. Kong,
Mod. Phys. Lett. A 14, 903 (1999); E.J. Chun, S.K. Kang, C.W. Kim,
U.W. Lee, Nucl. Phys. B 544, 89 (1999); K. Choi, K. Hwang,
E.J. Chun, hep-ph/9811363; A. Joshipura, S. Vempati,
hep-ph/9903435.

\bibitem{bilinear} A.Y. Smirnov, F. Vissani, Nucl. Phys. B 460, 37
(1996); E. Nardi, Phys. Rev. D 55, 5772 (1997); T. Banks, Y. Grossman,
E. Nardi, Y. Nir, Phys. Rev. D 52, 5319 (1995); R. Hempfling,
Nucl. Phys. B 478 , 3 (1996); H. Nilles, N. Polonsky, Nucl. Phys.
B 484, 33 (1997); C. Liu, Mod. Phys. Lett. A 12, 329 (1997);
B. Mukhopadhyaya, S. Roy, F. Vissani, Phys. Lett. B 443, 191
(1998); D.E. Kaplan, A. Nelson, hep-ph/9901254; J. Valle,
hep-ph/9712277.

\bibitem{sk} Super-Kamiokande Collaboration, Y. Fukuda {\em et al.},
Phys. Rev. Lett. 81, 1562 (1998). 

\bibitem{solar} J.N. Bahcall, M.H. Pinsonneault,
Rev. Mod. Phys. 67, 781 (1995); J.N. Bahcall, S. Basu,
M.H. Pinsonneault, Phys. Lett. B 433, 1 (1998).

\bibitem{moscow} V.M. Lobashov (for Troitsk Collaboration), Proc.~of
Neutrino '98, Takayama, Japan, June 1998.

\bibitem{mainz} C. Weinheimer (Mainz Collaboration), private
communication. 

\bibitem{hm} L. Baudis {\em et al.}, Phys. Rev. Lett. 83, 41 (1999);
H.V. Klapdor-Kleingrothaus, hep-ex/9901021, Proc.~Int.~Conf. on Lepton-
and Baryon Number Non-Conservation, Trento, Italy, April 1998,
IOP, Bristol, 1999.

\bibitem{review} For recent reviews, see
G. Bhattacharyya, hep-ph/9709395, Proc. BEYOND 97, Castle
Ringberg, Germany, June 1997, eds.~H.V. Klapdor-Kleingrothaus,
H. P\"{a}s, IOP, Bristol, 1998 (and its update to appear in the
Proc. of the Int.~WEIN98 Conf., Santa Fe, USA, June 1998);
Nucl. Phys. Proc. Suppl. 52A, 83 (1997);  
H. Dreiner, hep-ph/9707435;
R. Barbier {\it et al.}, hep-ph/9810232;
P. Roy, hep-ph/9712520. 

\bibitem{babu_moha} K.S. Babu, R.N. Mohapatra, Phys. Rev. Lett. 64,
1705 (1990); K. Enqvist, A. Masiero, A. Riotto, Nucl. Phys. B 373, 95
(1992); E. Roulet, D. Tommasini, Phys. Lett. B 256, 218 (1991). 

\bibitem{barbieri} R. Barbieri, M. Guzzo, A. Masiero, D. Tommasini,
Phys. Lett. B 252, 251 (1990).   

\bibitem{fuji} K. Fujikawa, R. Shrock, Phys. Rev. Lett. 45, 963
(1980).

\bibitem{ng} K.L. Ng, Z. Phys. C 48, 289 (1990). 

\bibitem{faessler} V. Bednyakov, A. Faessler, S. Kovalenko,
Phys. Lett. B 442, 203 (1998) 

\bibitem{genius} H.V. Klapdor-Kleingrothaus, Proc. BEYOND 97
\cite{review}; J. Hellmig, H.V. Klapdor-Kleingrothaus, Z. Phys. A 359,
351 (1997); H.V. Klapdor-Kleingrothaus, M. Hirsch, Z. Phys. A 359, 361
(1997); H.V. Klapdor-Kleingrothaus, J. Hellmig, M. Hirsch, J. Phys. G
24, 483 (1998); H.V. Klapdor-Kleingrothaus, Int. J. Mod. Phys. A 23,
3953 (1998).

\bibitem{bonn} J. Bonn {\em et al.}, Nucl. Instr. Meth. A 421, 256
(1999).

\bibitem{wein} Bhattacharyya in \cite{review} (to appear in the
Proc.~of the International WEIN98 Conference).

\bibitem{huitu} A. Faessler, T.S. Kosmas, S. Kovalenko,
J.D. Vergados, hep-ph/9904335; K. Huitu, J. Maalampi, M. Raidal,
A. Santamaria, Phys. Lett. B 430, 355 (1998).

\bibitem{d0} B. Abott {\em et al.}, D0 Collaboration, hep-ex/9907019.

\bibitem{hirsch} M. Hirsch, J. Valle, hep-ph/9812463; A. Faessler,
S. Kovalenko, F. Simkovic, Phys. Rev. D 58, 055004 (1998).

\bibitem{joshipura} Joshipura, Vempati in \cite{recent}.

\bibitem{beacom} J.F. Beacom, P. Vogel, hep-ph/9907383. 

\bibitem{sn} J.M. Lattimer, J. Cooperstein, Phys. Rev. Lett. 61, 23
(1988); R. Barbieri, R.N. Mohapatra, Phys. Rev. Lett. 61, 27
(1988). 

\bibitem{pdg} The Review of Particle Physics, C. Caso {\em et al.},
Eur. Phys. J. C 3, 1 (1998) (and 1999 off-year partial web-update). 

\bibitem{kim} J.E. Kim, Phys. Rev. Lett. 41, 360 (1978);
hep-ph/9904312.

\bibitem{gninenko} S.N. Gninenko, Phys. Lett. B 452, 414 (1999);
M. Maltoni, M.I. Vysotsky, hep-ph/9804464.  

\bibitem{munu} Y. Declais, private communication.

\bibitem{okun} A. Cisneros, Astrophys. Space Sci. 10, 87 (1971); 
L.B. Okun, M.B. Voloshin, M.I. Vysotsky, Sov. Phys. JETP 64, 446
(1986). For a pedagogic discussion, see R.N. Mohapatra,
P.B. Pal, ``Massive neutrinos in Physics and Astrophysics'' (World
Scientific, Singapore, Second Edition, 1998).

\bibitem{largemunu} M. Voloshin, Sov. J. Nucl. Phys. 48, 512 (1988);
R. Barbieri, R.N. Mohapatra, Phys. Lett. B 218, 225 (1989);
K.S. Babu, R.N. Mohapatra, Phys. Rev. Lett. 63, 228 (1989);
Phys. Rev. D 43, 2278 (1991); G. Ecker, W. Grimus, H. Neufeld,
Phys. Lett. B 232, 217 (1989); D. Chang, W. Keung, G. Senjanovic,
Phys. Rev. D 42, 1599 (1990); M. Leurer, N. Marcus, Phys. Lett. B
237, 81 (1990); J.C. Montero, V. Pleitez, hep-ph/9907212.

\bibitem{voloshin} Voloshin in \cite{largemunu}. 




\end{thebibliography}
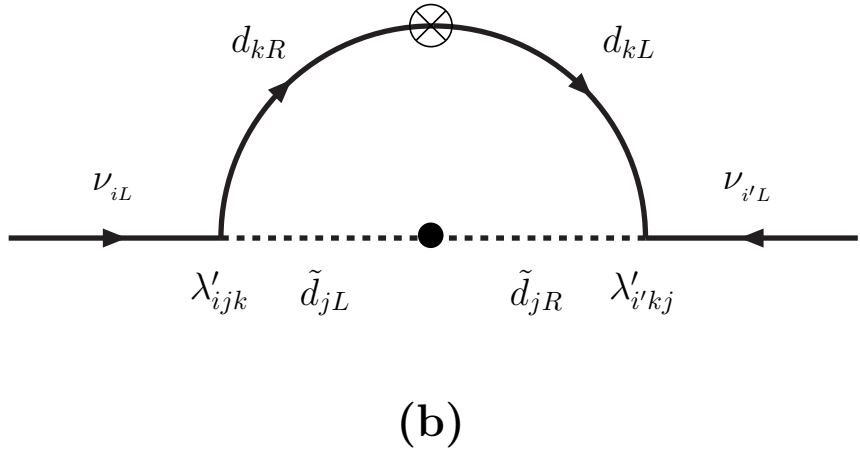
\end{document}